\documentstyle[aps,preprint]{revtex}

\tightenlines
\def\Ai{\hbox{\hbox{${\cal A}$}}\kern-1.9mm{\hbox{${/}$}}}
\def\Vi{\hbox{\hbox{${\cal V}$}}\kern-1.9mm{\hbox{${/}$}}}
\def\Di{\hbox{\hbox{${\cal D}$}}\kern-1.9mm{\hbox{${/}$}}}
\def\lam{\hbox{\hbox{${\lambda}$}}\kern-1.6mm{\hbox{${/}$}}}
\def\D{\hbox{\hbox{${D}$}}\kern-1.9mm{\hbox{${/}$}}}
\def\A{\hbox{\hbox{${A}$}}\kern-1.8mm{\hbox{${/}$}}}
\def\V{\hbox{\hbox{${V}$}}\kern-1.9mm{\hbox{${/}$}}}
\def\parz{\hbox{\hbox{${\partial}$}}\kern-1.7mm{\hbox{${/}$}}}
\def\B{\hbox{\hbox{${B}$}}\kern-1.7mm{\hbox{${/}$}}}
\def\R{\hbox{\hbox{${R}$}}\kern-1.7mm{\hbox{${/}$}}}
\def\si{\hbox{\hbox{${\xi}$}}\kern-1.7mm{\hbox{${/}$}}}

\begin{document}
\draft

\title{$U(1)\times SU(2)$ gauge theory of\\
 underdoped cuprate superconductors}
\author{P.A. Marchetti,$^a$  Zhao-Bin Su,$^b$ Lu Yu$^{c,b}$}

\address{ $^a$ Dipartimento di Fisica ``G. Galilei",
INFN, I--35131 Padova, Italy\\
$^b$Institute of Theoretical Physics, CAS, Beijing 100080, China\\
$^c$International Centre for Theoretical Physics, I-34100 Trieste, Italy}

\maketitle

\begin{abstract}
The $U(1)\times SU(2)$ Chern-Simons gauge theory is applied to study
the  2-D $t-J$ model describing the normal state of  underdoped
 cuprate superconductors. The $U(1)$ field produces a flux phase
for holons converting them into Dirac-like fermions, while the $SU(2)$
field, due
to the coupling to holons gives rise to a gap for spinons. An effective
low-energy action involving holons, spinons and a self-generated $U(1)$
gauge field is derived. The Fermi surface and electron spectral function
obtained are consistent with photoemission experiments. The theory
predicts  a minimal gap proportional to  doping concentration.
It also explains anomalous transport properties including linear
$T$ dependence of the in-plane resistivity.
\end{abstract}


\narrowtext

The proximity of superconductivity (SC) to antiferromagnetism (AF)
in reference compounds is a distinct feature of the high-$T_c$
superconductors.  Upon doping the AF goes away, giving rise to SC.
At the same time, the  Fermi surface (FS) is believed
to develop from small pockets around $(\pm {\pi \over 2}, \pm {\pi \over 2})$
\cite{shr}, anticipated for a doped Mott insulator,
  to a large one around $(\pi, \pi)$, expected from the electronic
structure calculations. To understand this crossover
is one of the key issues in resolving the high $T_c$ puzzle. For this reason,
the underdoped samples present particular interest due to the strong
interplay of SC with AF.
A "spin gap" or "pseudogap" has been invoked to explain the reduction of
magnetic susceptibility $\chi$ below certain characteristic temperature $T^*$
 and suppression of the specific heat compared with the linear
$T$ behavior\cite{mil}. This gap also shows up in
transport properties, neutron scattering, and NMR relaxation
rate. The angle-resolved photoemission spectroscopy (ARPES)
 data show clear Fermi level
 crossing in the (0, 0) to $(\pi, \pi)$ direction, but no such crossing was
detected in the $(0, \pi)$ to $(\pi, \pi)$ direction\cite{mar}.
The observed pseudogap
above $T_c$ is consistent with d-wave symmetry.
Theoretically there have been two competing approaches: One starting from the
Mott-Hubbard insulator, advocated by Anderson\cite{and} using the concept of
spin liquid, while  the other starting from the Fermi liquid (FL) point of
view.
Along the first line, P.A. Lee and his collaborators have consecutively
developed
the $U(1)$\cite{ln} and $SU(2)$\cite{wen} gauge theories, whereas the second
approach has been elaborated by Kampt and Schrieffer\cite{ks}, and Chubukov
and his collaborators\cite{chu}.

In this paper we apply to the $t-J$ model the $U(1)\times SU(2)$
Chern--Simons bosonization scheme for two-dimensional (2D)
fermion systems\cite{fro}.
This scheme provides a
decomposition of the electron field into a product of two ``semionic"
fields, advocated by R.B. Laughlin\cite{lau},
 one carrying the charge (holon) and the other carrying the spin (spinon).
 It has been shown in \cite{marc}
 that a mean-field  treatment of a dimensional
reduction of such bosonization procedure,
keeping the "semionic" nature of spinons and holons,
reproduces the exact results obtained by Bethe--Ansatz and Luttinger liquid
techniques, when applied to the 1D
$t-J$ model at $t \gg J$.  For  the underdoped 2D $t-J$ model we  neglect
 the feedback of holon fluctuations on the
$U(1)$  field $B$ and spinon fluctuations on the $SU(2)$  field
$V$. The holon
field is then a fermion and the spinon field a hard--core boson.
Within this approximation we show that the $B$ field  produces a
flux--phase for the holons, converting them into Dirac--like fermions;
the $V$ field, taking into account the feedback of holons
  produces a gap for the spinons vanishing in the
 zero doping limit, at  $(\pm {\pi\over 2}, \pm {\pi\over 2})$.
A low--energy effective action in terms of spinons, holons and a self
generated $U(1)$ gauge field is derived. Neglecting  the gauge fluctuations,
the holons are described by a FL with FS given by 4
"half-pockets" centered at  $(\pm {\pi\over 2}, \pm {\pi \over
2})$ and  one  reproduces the results for the electron
spectral function obtained  in mean field approximation (MFA)
\cite{dai}, in
qualitative agreement with the ARPES data \cite{mar}
 for underdoped cuprates.
Due to coupling to massless holons, gauge
fluctuations are \underbar{not} confining, but nevertheless yield an
attractive
interaction between spinons and holons leading to a bound state
in 2D with  electron quantum numbers.
The renormalisation effects due to gauge
fluctuations  induce non--FL behaviour for the
composite  electron, including the linear in $T$ resistivity
discussed earlier\cite{iw}.
This formalism describes  a smooth crossover upon doping from
the long range ordered (LRO) AF state to short ranged (disordered) AF state
with
a gap in the excitation spectrum.
The minimal  gap is proportional to the doping concentration and
the gap does not vanish in any direction.

much smaller

The euclidean action for the $t-J$ model in 2D can be represented in terms
of a fermionic spinless holon field $H$, coupled to a $U(1)$ gauge field
$B$, and a spin ${1\over 2}$ complex hard--core boson $\Sigma_\alpha, \alpha
=1, 2$, satisfying the constraint $\Sigma^*_\alpha \Sigma_\alpha =1$,
coupled to an $SU(2)$ gauge field $V$, and it is given by \cite{marc}:
\widetext
\begin{eqnarray}
& S = \int^\beta_0 d\tau {\displaystyle \sum_j}[ H^*_j \left(\partial_\tau
- i B_0 (j) -
\delta \right) H_j + i B_0 (j)
+ (1 - H^*_j H_j) \Sigma^*_{j\alpha} \left(\partial_\tau + i V_0 (j)
\right)_{\alpha\beta} \Sigma_{j\beta}] \nonumber \\
& +  {\displaystyle \sum_{\langle ij \rangle}}\left[ \left(-t H^*_i e^{- i
\int_{\langle ij \rangle} B} H_j \Sigma^*_{i\alpha} (P
e^{i \int_{\langle ij \rangle} V} )_{\alpha\beta} \Sigma_{j\beta} +
h.c.\right)\right. \nonumber \\
& \left. + {J\over 2} (1 - H^*_i H_i) (1 - H^*_j H_j) (|\Sigma^*_{i\alpha} (P
e^{i \int_{\langle ij \rangle} V})_{\alpha\beta} \Sigma_{j\beta}|^2 -
{1\over 2} ) \right]
-2 S_{c.s.} (B) + S_{c.s.} (V)
\label{act}\end{eqnarray}
\narrowtext where $P(\cdot)$ denotes the path--ordering,
$\delta$ the chemical potential for the dopants  and
$S_{c.s.} (B) = {1\over 4 \pi i} \int d^3 x \epsilon_{\mu\nu\rho}
B^\mu \partial^\nu B^\rho,$
$
S_{c.s.} (V) = {1\over 4\pi i} Tr \int d^3 x \epsilon_{\mu\nu\rho} (V^\mu
\partial^\nu V^\rho + {2\over 3} V^\mu V^\nu V^\rho)
$
are the Chern--Simons action for the gauge fields (we refer to \cite{marc} for
further details). The electron field at  site $j$ is
decomposed as\cite{fro,marc}:
$
c_{j\alpha} = e^{-i \int_{\gamma_j} B} H^*_j (P e^{i \int_{\gamma_j}
V})_{\alpha\beta} \Sigma_{j\beta},
$
where $\gamma_j$ is a straight line joining  site $j$ to $\infty$ in a
fixed time plane.

The (local) gauge invariances of (\ref{act}) are:
$ U(1):
H_j \rightarrow  H_j e^{i \Lambda_j}, B_\mu (x)
 \rightarrow B_\mu
(x) + \partial_\mu \Lambda (x), \; \Lambda (x)\in {\bf R};
$
$
SU(2): \Sigma_j \rightarrow R^\dagger_j \Sigma_j,
\;V_\mu (x) \rightarrow R^{\dagger} (x) V_\mu (x) R(x) + R^\dagger (x)
\partial_\mu
R(x), \; R(x) \in SU(2);
$
 and holon/spinon (h/s) gauge:
$
H_j \rightarrow H_j e^{i \xi j}, \; \Sigma_j \rightarrow e^{-i
\xi j} \Sigma_j, \; \xi_j \in {\bf R}.
$
 We gauge--fix the $U(1)$ symmetry  by
imposing Coulomb  condition for $B$. To retain the bipartite structure
induced  by
AF interactions, we gauge--fix the $SU(2)$ symmetry  by
a ``N\`{e}el gauge" condition:
$
\Sigma_j = \sigma_x^{|j|} (\matrix{ 1 \cr
0 }),\; |j| = j_1 + j_2.
$
Now we split the integration over the $V$ field into an integration over
a field $V^c$ satisfying the Coulomb condition,
$\sum^2_{\mu=1} \partial^\mu V_\mu^c =0 $
(from now on $\mu = 1,2)$, and its gauge transformations in terms of an
$SU(2)$--valued scalar field $g$.
Integrating  over $V_0$ and $B_0$, we obtain
\begin{eqnarray}
V^c_\mu = & {\displaystyle \sum_j} (1 - H^*_j H_j) (\sigma_x^{|j|} g^\dagger_j
{\sigma_a\over 2} g_j \sigma^{|j|}_x)_{11} \partial_\mu {\rm arg} \
(x-j) {\sigma_a\over 2}, \nonumber \\
B_\mu = & \bar B_\mu + \delta B_\mu, \quad \delta B_\mu (x)=
{1\over 2} {\displaystyle \sum_j} H^*_j H_j \partial_\mu {\rm arg} \
(x-j),\nonumber
\end{eqnarray}
where $e^{i \int_{\partial p}\bar B} = -1$ for every plaquette $p$,
and $\sigma_a$ are the Pauli matrices.

Following  the strategy   in  1D\cite{marc},
 we  write down the partition
function of holons in a  $g$ background  in terms of first
quantized Feynman path integral, and
find an {\it a priori} upper
bound on it. We then look for a holon-dependent configuration $g^m, V^c (g^m)$
saturating the bound,
taken as the starting point to add spinon
fluctuations.  This can be justified in the limit $ t \gg J$, because
the effective mass of holes is very heavy\cite{ly}.
 For an arbitrary given holon configuration
the term $(\sigma^{|i|}_x g^\dagger_i P (e^{i \int_{\langle ij \rangle} V}
) g_j
\sigma_x^{|j|})_{11}$  appears for a fixed link $\langle ij\rangle$
 either in the ``worldlines" of holons or in the Heisenberg term, but
never simultaneously; this permits a separate optimization of the
two cases (see \cite{marc}).
Using techniques  adapted from the proof of diamagnetic inequality
\cite{fro1},
assuming translational invariance for the  minimizing configuration $g^m$,
 neglecting the quartic pure holon term ($\delta \ll 1$)
in (\ref{act}), and making use of results of \cite{has},
it follows that for  the optimal
configuration (see \cite{marc1} for further details):
$V_{\mu}^c(g^m) =\bar V_\mu (g^m)+
\sum_j{(-1)^{|j|} \over 2} \partial_\mu
{\rm arg }\ (x-j) {\sigma_z \over 2},$
\begin{equation}
\bar V_\mu (g^m) = {\displaystyle \sum_j}  (-1) H^*_j H_j{(-1)^{|j|} \over
2} \partial_\mu
{\rm arg }
\ (x-j) {\sigma_z \over 2}
\label{vmu}\end{equation}
and on links belonging to the  holon worldlines
\begin{equation}
\left(\sigma_x^{|i|} g^{m \dagger}_i P ( e^
{i\int_{\langle ij \rangle}\bar V (g^m)})
 g^m_j \sigma_x^{|j|}\right)_{11} \sim 1,
\label{hop}\end{equation}
while on links in the Heisenberg term
\begin{equation}
\left(\sigma^{|i|}_x g^{m \dagger}_i P (e^{i\int_{\langle ij \rangle} \bar
V (g^m)})
g^m_j \sigma_x^{|j|}\right)_{11} \sim 0.
\label{heis}\end{equation}
Here $\bar V_\mu (g^m)$ is the slowly varying part of the $SU(2)$ gauge
field related
to holons, and the physical meaning of (\ref{hop}) and (\ref{heis})
will be explained later (after eq. (\ref{rpr})).  We  represent
$g_j= \exp{[-{ i \over 2}\sum_{l\neq j}(-1)^{|l|}\sigma_z{\rm arg} (j-l)]}
R_j \exp{[i{\pi \over 2} (-1)^{|j|}\sigma_y H^*_jH_j]}$, where
$R$ is  an $SU(2)$ -- valued field,  written in CP$^1$ form:
\begin{equation}
R_j = \left(\matrix{b_{j1} & - b^*_{j2} \cr
b_{j2} & b^*_{j1} \cr} \right) ,\;\;\; b^*_{j\alpha} b_{j\alpha} =1
\label{r}\end{equation}
(no summation over $j$), describing the spinon fluctuations around $g^m_j$
(for which
$R_j= \hat I$).
With suitable field redefinition, using the $SU(2)$ invariance of the
$g$ measure, the action of the $t-J$ model can be \underbar{exactly}
rewritten as $S=S_h + S_s$, where
\widetext
\begin{eqnarray}
& S_h  = \int^\beta_0 d\tau {\displaystyle \sum_j} H^*_j (\partial_\tau -
(\sigma_x^{|j|} R^\dagger_j \partial_\tau R_j \sigma_x^{|j|})_{11}
 -\delta) H_j\nonumber \\
& + {\displaystyle \sum_{\langle ij \rangle}} [- t H^*_i e^{-i\int_{\langle
ij \rangle} \bar B +\delta B} H_j
(\sigma_x^{|i|} R^\dagger_i(P e^{i\int_{\langle ij \rangle} \bar V + \delta
V}) R_j
\sigma_x^{|i|})_{11} + h.c. ]\nonumber \\
&S_s  = \int^\beta_0 d\tau {\displaystyle \sum_j}
(\sigma_x^{|j|} R^\dagger_j \partial_\tau R_j \sigma_x^{|j|})_{11}\nonumber \\
&+ {\displaystyle \sum_{\langle ij \rangle}} {J \over 2} (1 - H^*_i H_i) (1
- H^*_j H_j)
\{|(\sigma_x^{|i|} R^\dagger_i(P e^{i\int_{\langle ij \rangle} \bar V +
\delta V}) R_j
\sigma_x^{|j|})_{11}
|^2 - {1\over 2} \}.
\label{act1}\end{eqnarray}
\narrowtext
Notice that for small hole concentration $(\delta
\ll 1)$, $\bar V$ is a slowly  varying field.


We now make the first approximation: suppose  in (\ref{act1}) the
fluctuations of the $V$ field, due to the spinon
fluctuations $R$ are small enough that  we can set $\delta V =0$.
Since the main effect of these fluctuations is to convert
the $SU(2)$ gauge invariant spinon field
 into a semion field(see \cite{fro}), to be consistent, we neglect also the
feedback of the holon field on $B$ responsible for its semion nature,
 i.e.  set $\delta B=0$ as well.
We believe that the proper account of the
statistics of gauge--invariant spinon and holon fields is
less crucial in 2D than in 1D, as we will see later.
Let us consider the variable $R^\dagger_i P(^{i\int_{\langle ij
\rangle}\bar V} ) R_j = $
\begin{equation}
\left(\matrix{ \alpha b^*_{i1} b_{j1} +
\alpha^*  b^*_{i2} b_{j2} & - \alpha b^*_{i1}
b^*_{j2} + \alpha^*  b^*_{i2} b^*_{j1} \cr
- \alpha  b_{i2} b_{j1} + \alpha^*  b_{i1} b_{j2}
 & \alpha b_{i2} b^*_{j2} + \alpha^*  b_{i1} b^*_{j1} \cr}
\right),
\label{rpr}\end{equation}
where $\alpha = \exp ({{i \over 2}\int_{\langle ij \rangle} \bar V_z})$.
 In the hopping term of holons only the diagonal elements of
(\ref{rpr}) appear, a kind of gauge invariant Affleck-Marston (AM)
variable \cite{am}; in
the Heisenberg  term only the off--diagonal elements appear, a kind of
gauge invariant resonant valence bond (RVB) variable.
According to the minimization arguments given above,
 the mean value of the AM gauge variable is  $s$-like,
real and close to 1 (see eq.(\ref{hop})), while the RVB order parameter
should be rather small (see eq.(\ref{heis})).
using a
We now obtain a low--energy continuum action for spinons rescaling
the model to  a lattice spacing $\varepsilon \ll 1$ and neglecting higher
order terms in $\varepsilon$. As it is standard in AF systems
we define $\vec n_j =
b^*_{j\alpha} \vec \sigma_{\alpha\beta} b_{j\beta}$ and assume\cite{fra}:
$
\vec n_j \sim \vec \Omega_j + (-1)^{|j|} \varepsilon \vec L_j,\;
\vec \Omega_j^2 = f \leq 1, \;
$
(j)) +
More precisely  the fields $\vec \Omega, \vec L$ are defined  on a
sublattice, e.g.
$\vec\Omega_j \equiv \vec\Omega_{j_1+ {1\over 2}, j_2}, \vec L_j \equiv
\vec L_{j_1 + {1\over 2}, j_2}, j_1 = j_2  \ {\rm mod (2)} $
and they describe the AF and  ferromagnetic fluctuations, respectively.
It is  useful to write $\vec\Omega$ in CP$^1$ form:
$\vec\Omega = z^*_\alpha \vec\sigma_{\alpha\beta} z_\beta,$
with  $z_\alpha , \alpha = 1, 2$,  a spin ${1\over 2}$ complex hard--core
boson satisfying the constraint
$z^*_\alpha z_\alpha = f.$

  Evaluating the holon contribution  in MFA, using the absence
of the  ``$\theta"$ term  in 2D\cite{fra}, and integrating out
over $\vec L$, we obtain a low--energy continuum limit NL$\sigma$ model with
action
\begin{equation}
S_s = \int d^3 x {1\over g} [(\partial_0 \vec \Omega)^2 + v^2_s
(\partial_\mu \vec \Omega)^2 + \vec\Omega^2 \bar V^2_z ] - {1\over g}
(\Omega_z)^2 \bar V^2_z,
\label{sigma1}\end{equation}
where $g$ and $v_s$  are easily derived in terms of $J, t, \delta,
\varepsilon$.

 To consider the effect of the  $\bar V$ field, we replace the NL$\sigma$
constraint $\vec\Omega^2 = f$ by a softened version adding to the
lagrangian  a term $\lambda (\vec\Omega^2 - f)^2$,
and substitute $\bar V^2_z$
 by its statistical averaging over holon configurations, $\langle \bar
V^2_z \rangle$. Using a
sine--Gordon transformation \cite{edw}, we obtain $\langle \bar V^2_z
\rangle \sim  - \delta \ln
\delta$ (see \cite{marc1} for details). For  small $J$, the coupling
constant $g$
is small and the system is
in the ordered phase; the renormalization group flow in the absence of
perturbation drives $g_{eff}$ for large distances towards  its critical value;
the
mass perturbation induced by $\langle  \bar V^2_z \rangle$ should then
drive the system
from the ordered to the disordered phase (with short range order only).
(One might speculate that if we
 treat the holons as slowly
moving impurities, consistent with the known results in
the limit $t \gg J$\cite{ly}, this would lead to a kind of ``Anderson
localization"
considered in \cite{esc}).
Hence our system should
exhibit a mass gap $m (\delta)$ vanishing as $\delta \rightarrow 0$, and
absence of AFLRO (at least for $\delta \ll 1$, but  sufficiently big).
This provides a smooth crossover to the insulating AF regime.
The subleading
perturbation appearing in eq.(\ref{sigma1}) gives rise to a remnant spin-space
uniaxial AF interaction in short-ranged AF state.

We can summarize the above discussion by rewriting the NL$\sigma$ model action
in CP$^1$ form, neglecting the short range interactions, as
\begin{equation}
S^\star_s = \int d^3 x {1\over g} \left[|(\partial_0 - A_0)
z_\alpha|^2 + v_s^2 |(\partial_\mu + A_{\mu})
 z_\alpha|^2  + m^2 z^*_\alpha z_\alpha \right]
\label{cp1}\end{equation}

In the NL$\sigma$ model without mass term  ($\delta=0$), the
constraint $z^*_\alpha z_\alpha =f$ and the symmetry breaking condition,
e.g. $\langle z_1  \rangle \not = 0$, lead to excitations described by a
complex massless
field $S \equiv \langle z^*_1  \rangle z_2$,
with  relativistic massless dispersion relation
corresponding to the spin waves. In the NL$\sigma$ model with
mass term the absence of symmetry breaking and the effective softening of
the constraint lead to excitations described by the spin ${1\over 2}$
two--component complex field $z_\alpha$, with
massive dispersion
relations. However, the self--generated gauge field $A_{\mu}=z^*_\alpha
\partial_\mu z_\alpha$  confines the spin ${1\over 2}$
degrees of freedom and the actual excitations are described by a composite spin
1  spin--wave field.
As we shall see, the coupling to holons will induce
deconfinement of spin ${1\over 2}$ excitations.
In terms
of  fields $b_\alpha$, ( slave fermion approach),
one realises that the $z-$field in the reduced Brillouin
zone with two complex components  corresponds (at $\delta = 0$)
to  appearance of an  $s+id$ RVB
order parameter with a vanishing gap, in MFA, at four
points $(\pm {\pi \over 2},\pm {\pi \over 2}$) \cite{yos}. The NL$\sigma$
model for spinons can  also be derived in this representation. The low--energy
excitations of this model are fluctuations around these
points turned into massive ones by the $\langle \bar V^2_z \rangle$ term.
Physically, this is a coexisting $\pi$ flux plus $s+id$ RVB state. From our
estimate
 we expect the RVB $s+ id$ order parameter (\ref{heis})  to be much smaller
then the
AM order parameter (\ref{hop}).
These features are clearly shown in the numerical MFA
calculations of \cite{sheng}.


Now turn to  holons. We use a $U(1)$ gauge with
$e^{i\int_{\langle ij \rangle} \bar B}$ being purely imaginary and
assume spinons are in  disordered phase  with  AM parameter
$\sim 1$.
In the rescaled  $\varepsilon$ lattice,
 neglecting higher order terms in $\varepsilon$ and  $b$, the effective
action describes the usual 2--component Dirac (``staggered") fermions of the
flux phase\cite{dm},  with vertices of the
double--cone dispersion relations in the reduced Brillouin zone
centered at $(\pm {\pi \over 2}, \pm
{\pi \over 2})$ (in the $\varepsilon =1$ lattice),  with chemical
potential $\delta$. 

Define the four sublattices:  (1) for $j_1, j_2$ even, (2) for  $j_1$ odd
$j_2$ even, (3) for $j_1$ even $j_2$ odd, (4) for $j_1, j_2$ odd; they can be
grouped into two ``N\`{e}el sublattices"
$A = \{(1), (4) \}, B = \{(2), (3) \}.$ The holon field restricted to the
sublattice $(\#)$ is denoted by $H^{(\#)}$. Set:
\begin{eqnarray}
&\Psi^{(1)} \equiv \left(\matrix{\Psi^{(1)}_A \cr \Psi^{(1)}_B \cr} \right)
\equiv \left(\matrix{e^{-i {\pi\over 4}} H^{(1)} + e^{i{\pi\over 4}} H^{(4)}
\cr e^{-i{\pi\over 4}} H^{(3)} + e^{i{\pi\over 4}} H^{(2)} \cr} \right)
\nonumber\\
&\Psi^{(2)} \equiv \left(\matrix{\Psi^{(2)}_B \cr \Psi^{(2)}_A \cr} \right)
\equiv \left(\matrix{ e^{-i{\pi\over 4}} H^{(2)} + e^{i{\pi\over 4}}
H^{(3)} \cr
e^{- i{\pi\over 4}} H^{(4)} + e^{i{\pi\over 4}} H^{(1))} \cr} \right)
\nonumber \\
&\gamma_0 = \sigma_z, \gamma_\mu = (\sigma_y, \sigma_x), \A \equiv
\gamma_\mu A^\mu, \parz \equiv \gamma_\mu\partial_\mu,
\bar\Psi^{(\#)} = \gamma^0 \Psi^{(\#)\dagger}\nonumber
\end{eqnarray}
and assign charge $e_A = +1 (e_B =-1)$ to the fields on the $A (B)$
sublattice. Then, neglecting short range interactions,
 we obtain the low--energy continuum action for holons:
$$
S^\star_h = \int d^3 x \sum^2_{r=1}
\bar\Psi^{(r)} \Bigl(\gamma_0 (\partial_0 -
\delta - e^{(r)}A_0) + t (\parz - e^{(r)}
\A ) \Bigr) \Psi^{(r)}
$$
Here  $A_\mu$ is nothing but  the gauge field
for the h/s gauge.
We can use $S^\star = S^\star_s +S^\star_h $
 to compute, as in \cite{il}, the gauge field propagator induced
by the spinon and holon vacuum polarisation.
Since the spinon is massive, the corresponding vacuum polarization
would be  Maxwell--like.
Hence, in the absence of holons, it would logarithmically confine the
spinons. However,  excitations represented by $\Psi^{(1)}_B$ and
$\Psi^{(2)}_A$ describe a FL with  a small FS
 $(\varepsilon_F \sim O (\delta t)  )$
in the reduced Brillouin zone around the points $({\pi \over 2}, \pm
{\pi \over 2})$. Thus the vacuum polarization exhibits the Reizer
singularity \cite{ln} and the full gauge interaction is \underbar{not}
confining.
Nevertheless, since we are in 2D, the attractive force  mediated by the
gauge field is expected  to produce  bound states neutral
w.r.t. the h/s gauge, i.e. bound states with quantum numbers of the
spin wave (for a rigorous discussion of a similar problem,
see \cite{fro2}) and the electron\cite{chen}, respectively.
For this reason, neglecting "semion" nature can be justified
to some extent in 2D. Even if we neglect the gauge fluctuations, the existence
of  two bands  in the reduced Brillouin zone gives rise to a
"shadow band" effect (the spectral weight for the part of the pocket facing
$(\pi, \pi)$ is greatly reduced) due to
the presence of $\gamma-$matrices, leading to mixing of fermions of
these  bands. The situation is similar to the slave boson case \cite{dai}.

To conclude we summarize the main differences between 1D and 2D cases. In $1D$:
the NL$\sigma$ model of spinons contains
a $\theta=\pi$ topological term, yielding deconfinement;
absence of $\bar V$ term  makes the spinons massless;
there is no $\bar B$ term and there is only one holon band;
the h/s gauge field $A$  vanishes, hence
there is no attractive gauge force between holons and spinons, so their
statistics appears to be crucial.
In contrast, in 2D, without taking into account the "semion" nature
of holons and spinons, but considering the feedback of holons on
the $SU(2)$ gauge field, producing the spinon gap, we can already
understand quite a number of peculiar properties for underdoped cuprates:
normal state pseudogap, small FS, shadow bands, etc. The main features
of the MF calculation
\cite{dai} survive gauge fluctuations. Further consideration of
these fluctuations between holons and spinons provides a binding force
between them, and this composite electron shows non-FL behavior,
like linear $T$ dependence of resisitivity\cite{iw}, and others.
More detailed consideration of various physical properties within the
present model will be given elsewhere\cite{marc1}.

One of us (P.M.) would like to thank J. Fr\"ohlich and F. Toigo,
while L.Y. would like to thank A. Tsvelik, for
stimulating discussions. The work of P.M. was partially supported by
TMR Programme ERBFMRX-CT96-0045, whereas the work in Trieste benefitted
from the contract ERBFMRX-CT94-0438.


\end{document}